# COMPLEXITY EVALUATION OF NETWORK CONFIGURATIONS AND ABSTRACTIONS

Jose Moreno, Microsoft

*ABSTRACT*

*Computer networks have been traditionally configured by humans using command-line interfaces. Some network abstractions have emerged in the last 10 years, but there is no easy way of comparing them to each other objectively. Therefore, there is no consensus in the industry of what direction should modern network abstractions take, and the adoption of these abstractions lags.*

*In this paper I propose a comparison framework using metrics derived from graph structures to evaluate the simplicity, efficiency, and effectiveness of different network abstraction models.*

*The result of this comparison is that while some of the existing network abstractions are quite efficient to store network policy (such as the Kubernetes or the Cisco Application Centric Infrastructure models), others (notably public cloud) are still very infrastructure-centric and suffer from excessive complexity.*

## 1. INTRODUCTION

Computer networks are the basis of today's internetworked world. Not only they provide the basis upon which users connect to applications, but distributed application components in microservices architectures also rely more and more on the underlying networks to provide their functionality. Therefore, the stability of computer networks is of crucial importance for the availability of applications, as recent outages have proven [1]. According to [2], networking-related problems have been the single biggest cause of all IT service downtime incidents over the previous three years.

Some network outages are caused by failing components, but many of the outages are still due to misconfigurations and human error[3]. The risk of misconfigurations increases as networks grow more complex, and yet there are very few studies on the complexity of computer networks.[4]is one of the first studies on this topic, albeit focusing mostly on the complexity of the underlying network technologies such as routing protocols, but not on the logical models and interfaces used to configure them.

One way of reducing the human error component would be by automating every single network change. The meaning of the word "automating" here can be deceiving, since even if networks were configured with a pure Infrastructure-as-Code approach[5], the actual "code" (the policy) would be written and approved by humans.

A simplified network abstraction that reduces the complexity of network policy would decrease the potential for human error in changes (regardless of whether changes are carried out manually or in an automated way) and allow to check the syntax of network changes to automatically detect inconsistencies before they provoke a network outage. The computer network industry has been very sceptical when new abstractions have been introduced (see for example [6] for a subjective critic on the Cisco Application Centric Infrastructure object model), and adoption has usually been slow: the industry needs an unbiased criteria to evaluate the complexity and effectiveness of new abstractions proposed by actors such as network vendors, cloud providers or orchestration frameworks.

Many studies exist that evaluate the complexity of systems. Proven approaches([7],[8])consist of modelling the system into a graph and evaluating its complexity. And yet, the question about the definition of complexity stands. Different trains of thought exist in the literature to evaluate graph complexity:

- Statistical complexity (Shannon entropy): the complexity of a graph is equated to the information contained in it, or in other words, the size of the lossless compression of the graph[9].
- Statistic parameters of the graph: properties of graphs such as assortativity[10] and nestedness[11]can be used to evaluate its randomness.
- Algorithmic complexity (Kolmogorov complexity or K-Complexity [12]): some graphs will contain a lot of information (entropy), but that information might not be completely random. Instead, it would be possible to generate the same information by an algorithm. The Kolmogorov complexity describes the minimum length of a theoretical algorithm that would be able to generate a given graph in a Turing-complete machine. Even if the Kolmogorov-complexity cannot be calculated exactly, upper bounds can be determined [9].

And yet none of these complexities helps us, at least as long as networks are managed by humans. The Shannon entropy and measures such as assortativity and nestedness measure the randomness of a graph, but not its complexity from the perspective of the human operator of a computer network. The K-complexity is closer to this human-perceived complexity, but it is still not a good objective measure: even if a policy can be generated by an algorithm, it is not necessarily "simple" to understand, let alone to manage, for a human. For example, consider a sequence of numbers with the decimal values of an irrational number such as $2^{1/2}$:

- The numbers appear to be random, so the information contained is maximum according to the Shannon theorem.
- An algorithm can be devised that generates these apparently random numbers, so upper bounds for the Kolmogorov complexity can be found.
- Even if the algorithmic complexity is deemed to have upper bounds, for a human these numbers still appear as random. It wouldn't be possible for a human to "manage" that number sequence.

The reason why the previously mentioned approaches do not yield appropriate results to evaluate the complexity of computer networks is because not all objects in network abstractions contribute to complexity: while some are required to express the intended policy, others just contribute noise, and model parts of the network that are not relevant for the desired state. The more non-policy objects a network abstraction has, the more complex it will be to maintain. Hence, the vertices or nodes in the network graphs evaluated in this article will be separated into policy-related (p-vertices) and other types (infrastructure or i-vertices, and IP addresses or a-vertices).

The same way that there are different types of vertices in a network model, not all edges are the same. Some edges are loose couplings, meaning that if vertex A is related to vertex B, the network operator needs to type the name of vertex B to create the link. An example of loose couplings is the association of an access-control list to a router's interface, or the label-based model in Kubernetes. Other abstractions rather focus on tight couplings, where the vertex A and B need to exist before the operator can relate them, and a vertex cannot be deleted until all its edges are gone too (otherwise the edges would be invalid). For example, Cisco Application Centric Infrastructure and most public cloud network models follow this approach.

Therefore, these measures will be taken from each of the networks compared in the article:

- Number of nodes divided by the number of endpoints (physical hosts, virtual machines or Kubernetes pods) in the network, to get a first approximation of the size of the model.
- Total number of edges.
    - Number of loose-coupling edges
    - Number of tight-coupling edges
- Different types of nodes.
    - Policy-related node types
    - Infrastructure-related node types
- Excess degree of IP address nodes, to get a measure of the dependency of the model on the actual infrastructure.

This framework to compare network topologies and abstractions will be used in this article for two use cases:

1. To compare different topologies while using the same network abstraction model.
2. To compare the same topology modelled with different network abstractions.

## 2. COMPARING DIFFERENT TOPOLOGIES ON THE SAME ABSTRACTION

### 2.1. Topology 1: Complex Routing

The first use case for the proposed complexity evaluation framework is comparing the complexity of different topologies modelled with the same network abstraction. Three sample topologies in a public cloud (taking Microsoft Azure, although other public cloud network abstractions are very similar) have been chosen, which are designs that are often seen in workload deployments as Infrastructure-as-a-Service. All three topologies have the same traffic filtering requirements of two 4-tier applications plus a range of 4 shared services and 24 virtual machines in total, but implemented in different ways:

1. Topology 1: all traffic is filtered in a centralized network firewall.
2. Topology 2: part of the traffic is filtered in a centralized network firewall, the rest in distributed Access Control Lists (called Network Security Groups in Azure).
3. Topology 3: all of the traffic is filtered with distributed segmentation controls (Network Security Groups), and no centralized firewall is required.

Public cloud providers are interesting for this study because they offer a single API that models all aspects of the network: firewall rules, Layer-4 ACLs, routing, subnets, etc., which facilitates absorbing the model of the generated network and converting it to a graph. Hence, I will use a public cloud provider network abstraction (Microsoft Azure) as a baseline for these comparisons (other cloud providers have similar models for their network implementations).

The first topology to be analysed is going to have an unnecessary level of complexity, derived from a functional requirement: sending all inter-subnet traffic through a centralized firewall, even for subnets in the same virtual network. In this particular public cloud, this apparently simple requirement will derive in an elevated number of vertices in the resulting network model.

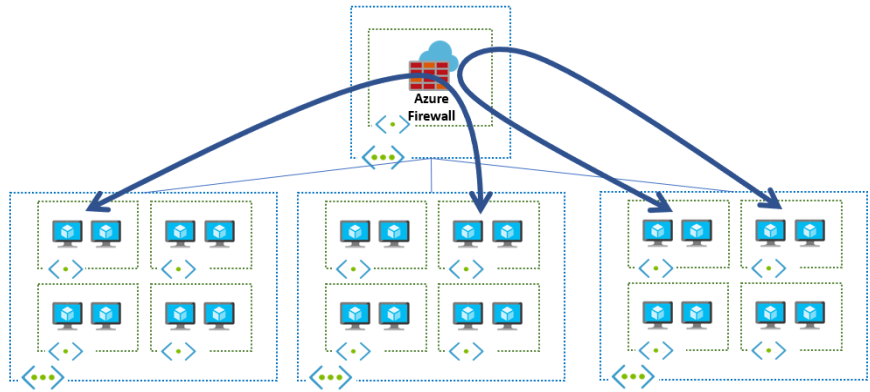

Figure 1. Topology 1 (Azure) - Increased routing complexity

To generate the model, an undirected graph has been generated with the Python library network x for further analysis and representations. This graph can be represented with different colours according to their type, yielding the picture below. The reader should not try to understand every vertex or edge of this graph, but just to get an idea about the graph dimensions:

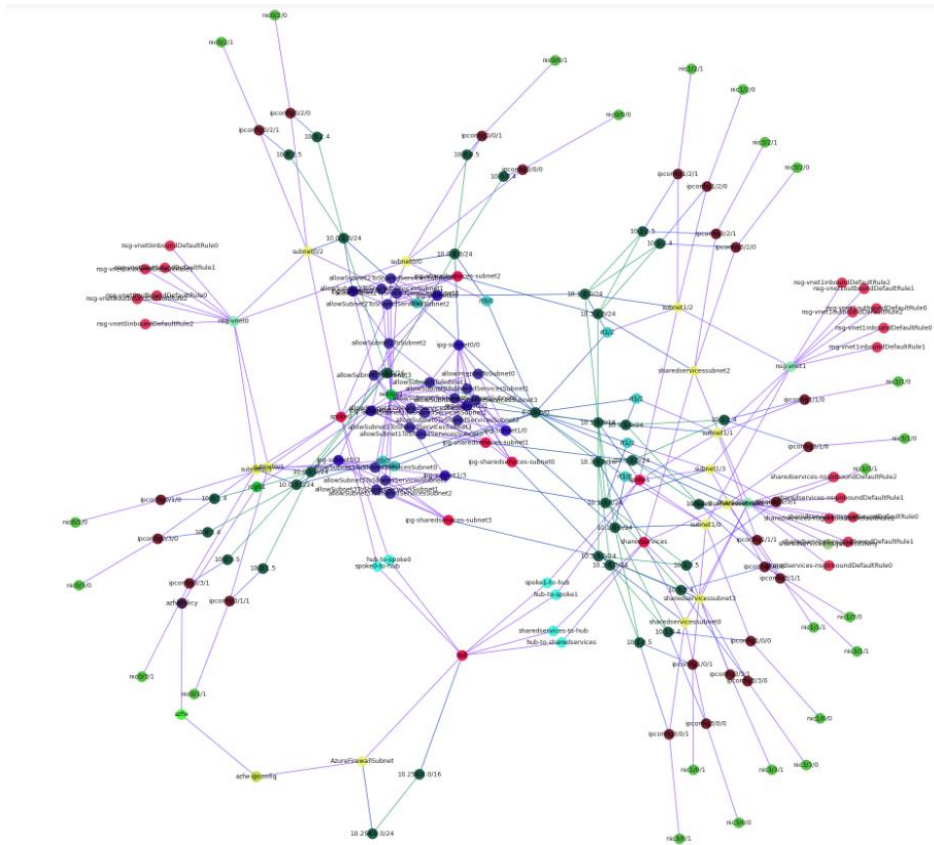

Figure 2. Topology 1 (Azure) – Graph representation

A first metric that can be derived is the number of nodes in the model by the number of endpoints $N_E$ (an endpoint is a virtual machine in public cloud). In this network there are 24 virtual machines (3 virtual networks or VNets, 4 subnets/VNet, 2 VMs/subnet.

The previous figure is difficult to comprehend. Instead, an alternative visualization providing a summary of the vertices per type is easier to interpret. The following representations shows edge aggregates in different colours (loose couplings in blue, tight couplings in blue, IP assignments in grey) and different thickness levels depending on their quantity. Vertices have as well a size depending on their quantity:

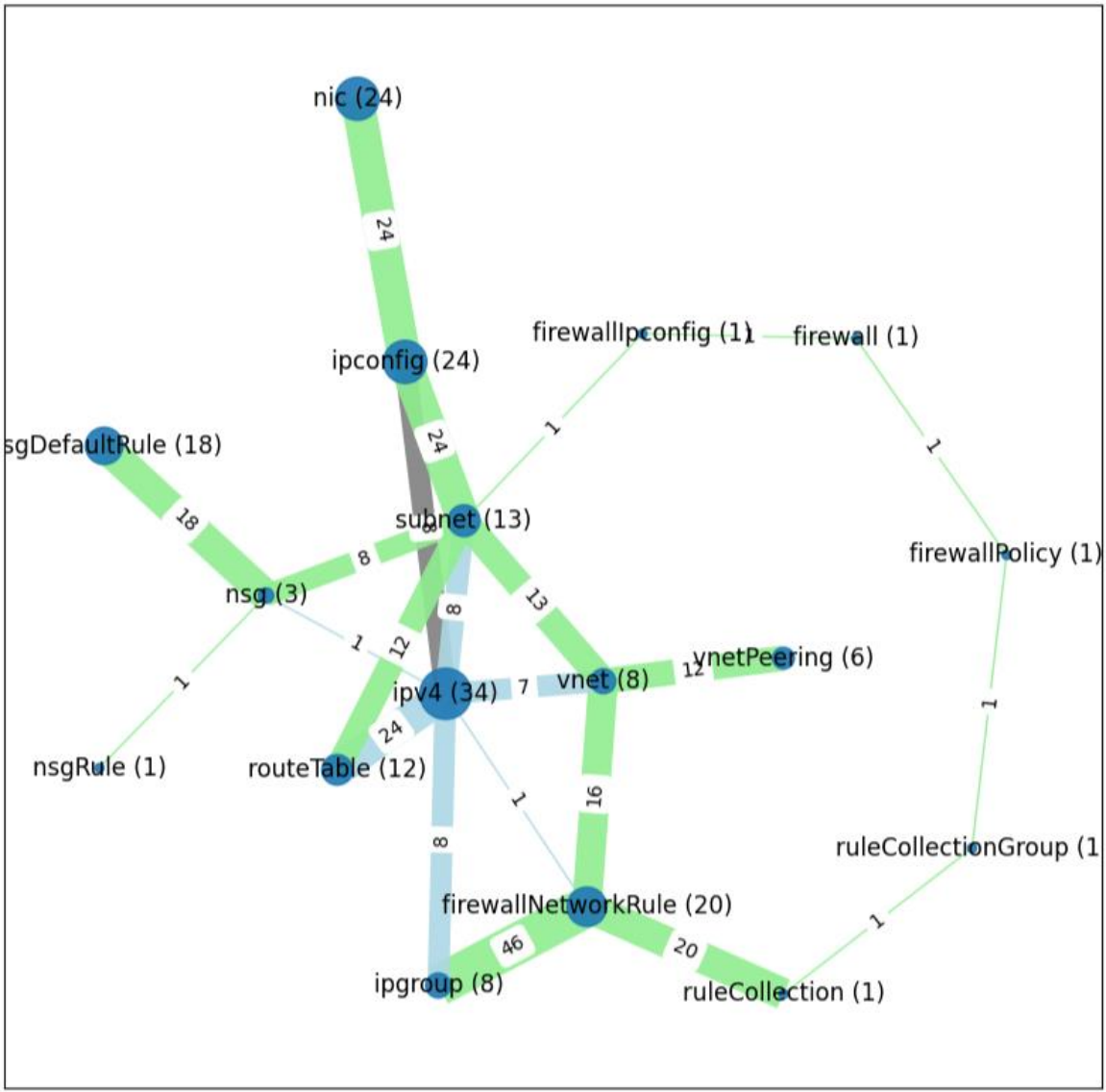

Figure 3. Topology 1 (Azure) – Vertex-type summary graph

The first conclusion that can be extracted out of this node-type-centric representation is that some of these vertex types contribute to express policy, but others don't:

- Network policy is expressed by vertices such as firewallPolicies, ruleCollections, or nsgRules. In the rest of the article, the node types that contain network policy information are called P-Types.
- Other vertex types such as subnets, vnets or ipconfigs do not contain any network policy, instead they are relevant for the underlying infrastructure. In the rest of the article, the node types that refer to the infrastructure, and do not contain network policy information, are called I-Types.

There are 34 vertices of the type "ipv4", which correspond to either individual IP addresses or IP prefixes for subnets and virtual networks, which are referenced with loose couplings by many different other resources, such as:

- Virtual networks
- Subnets
- IP Groups
- Routes
- IP configurations (and hence indirectly by network interfaces)

These loose couplings are a significant source for misconfigurations: since they need to be specified by the operator, there is a non-negligible risk of configuration errors or typos. Ideally, literal values should only have to be configured once in any network abstraction: for example, once the network administrator defines that a virtual network has a certain prefix, everything else that depends on that virtual network should reference the virtual network with its name or ID, and not with its IP address.

For example, in some designs you need to configure very specific static routes that match the value of some prefix somewhere else in the network (in Azure networking multiple instances of this situation can be found, such as the static routes that need to be configured in the gateway subnet, more information about static routes can be found in [13]). This means that every time that the network administrator changes a network prefix somewhere in the network, they need to change it somewhere else too. Failing to do so will result in a network inconsistency, and very likely in a network outage.

If we want to evaluate this factor of excessive reliance on IP addresses and prefixes for network policy, we can calculate the excess degree of the IP address vertices, excluding the "contains" edges ("contains" edges in the created graph relate subnets with the supernets containing them). This metric can be defined as follows:

$$\sum_{i=0}^{n}(d(x_i) - 1)$$

where $x_i$ is a literal vertex (such as an IP address typed manually) and $d(x_i)$ is the degree of that vertex. In other words, this is a measure of how often (more than once) an operator needs to type the same IP address so that the intended configuration is effective.

In this topology example, the total numbers of edges representing loose couplings to ipv4 nodes exceeding 1 is 37, which is part of the set of metrics that evaluate this topology.

All the previous data is summarized in this table:

Table 1. Topology 1 (Azure) – Complexity metrics

| Topology | Nodes/$N_E$ | L-Edges | T-Edges | I-Types | P-Types | IP-ED |
|---|---|---|---|---|---|---|
| Topology 1 (Azure) | 7.33 | 55 | 203 | 8 | 8 | 37 |

These values do not have any intrinsic meaning by themselves but will be useful when compared to other topologies.

## 2.2. Topology 2: Simplified Routing

The reader familiar with Azure might have realized that the example topology I used in the previous example has an unusual number of route tables, and the reason is because we imposed an artificial requirement that mandated it (sending intra-virtual-network traffic to a centralized firewall). Removing that requirement would decrease the number of route tables required, and hence the total amount of vertices that the network model would have:

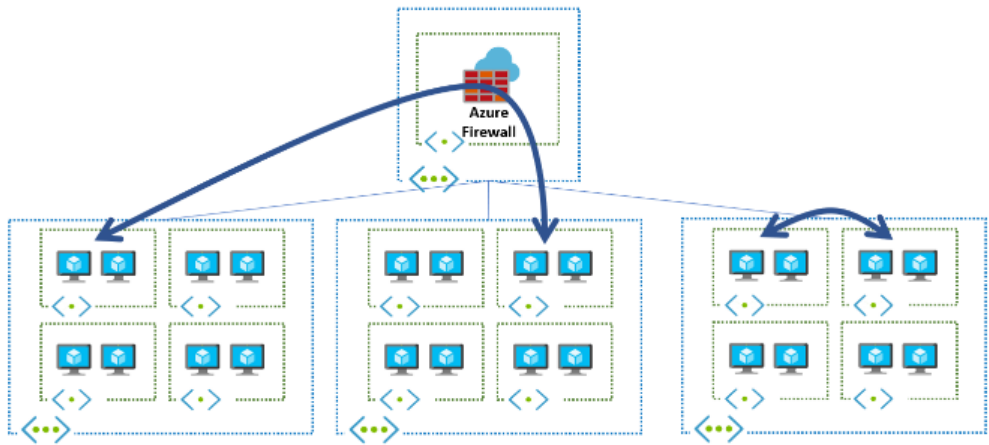

Figure 4. Topology 2 (Azure) – Simplified routing

A simplified representation of the graph summarizing the vertices by their type will show the consequence of simplifying the design:

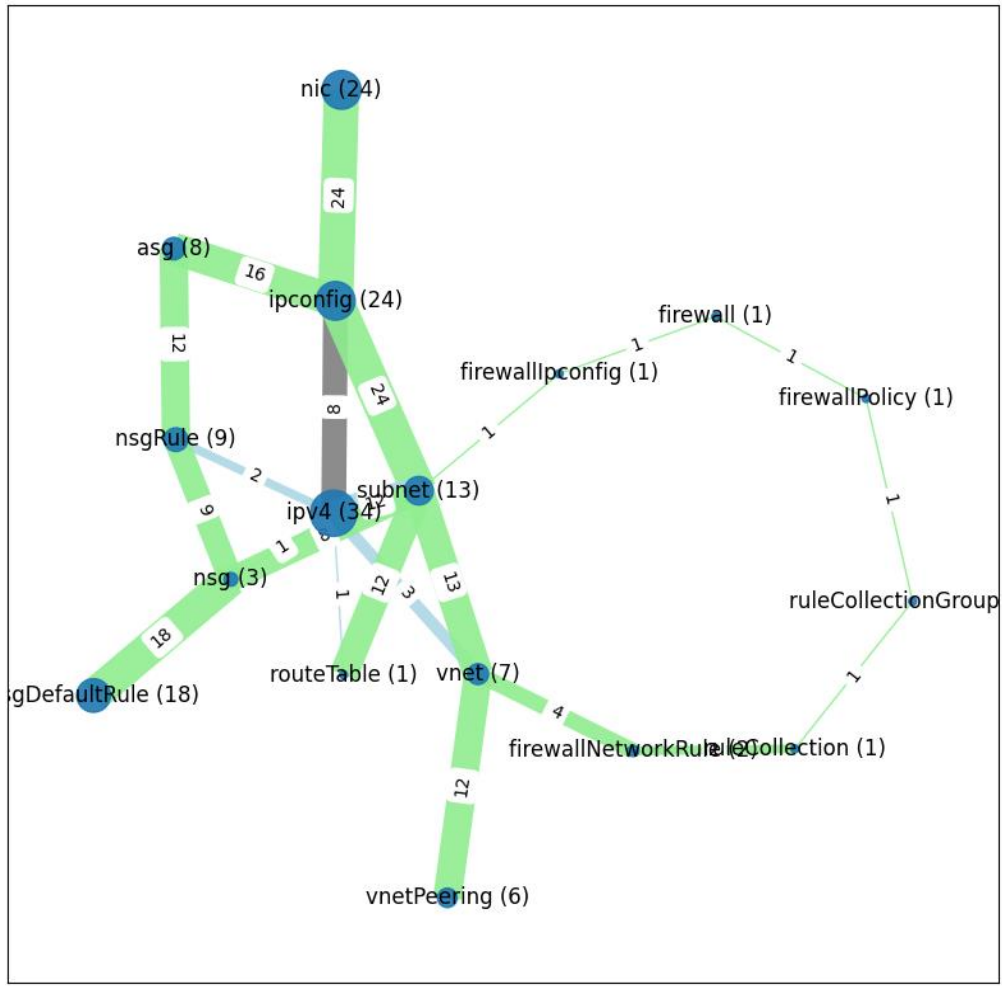

Figure 5. Topology 2 (Azure) – Vertex-type summary graph

It can be seen that the IPv4 nodes, a considerable source of complexity identified in the previous topology, have now fewer edges going to other nodes, specifically to route tables. Additionally, most of the edges are now tight couplings (and not loose couplings).

This decrease in the excess degree of IP addresses means that there are many fewer instances when the admin must statically configure the same value (an IP address) in two different places. For example, when defining the virtual network address space and a static route. Still, in our topology there are ranges that need to be configured more than once: the excess degree of IP address vertices is non-zero. For example, the neighbours for the IP address of one of our virtual networks has been configured manually as well in an IP group used in firewall rules, which is again an example of inefficiency.

Table 2. Topologies 1 and 2 (Azure) – Complexity metrics

| **Topology** | **Nodes/$N_E$** | **L-Edges** | **T-Edges** | **I-Types** | **P-Types** | **IP-ED** |
|---|---|---|---|---|---|---|
| Topology 1 (Azure) | 7.33 | 55 | 203 | 8 | 8 | 37 |
| Topology 2 (Azure) | 6.42 | 24 | 163 | 8 | 8 | 6 |

The network abstraction is the same (same number of infrastructure and policy vertex types I-Types and P-Types), and the overall number of nodes per endpointis significantly lower. However, the most significant reduction of complexity of this topology is the reduced number of loose couplings (L-Edges) with IP addresses, which results in a much lower excess-degree of IPv4-type nodes (IP-ED). In other words, operators need to type IP addresses much less frequently in this design.

### 2.3. Topology 3: Simplified Segmentation

We can further reduce the complexity of this network design. In the previous section, routing was simplified by leveraging the same route table for all subnets. Eliminating the centralized firewall so that segmentation is only performed by Network Security Groups will further decrease the complexity of the topology:

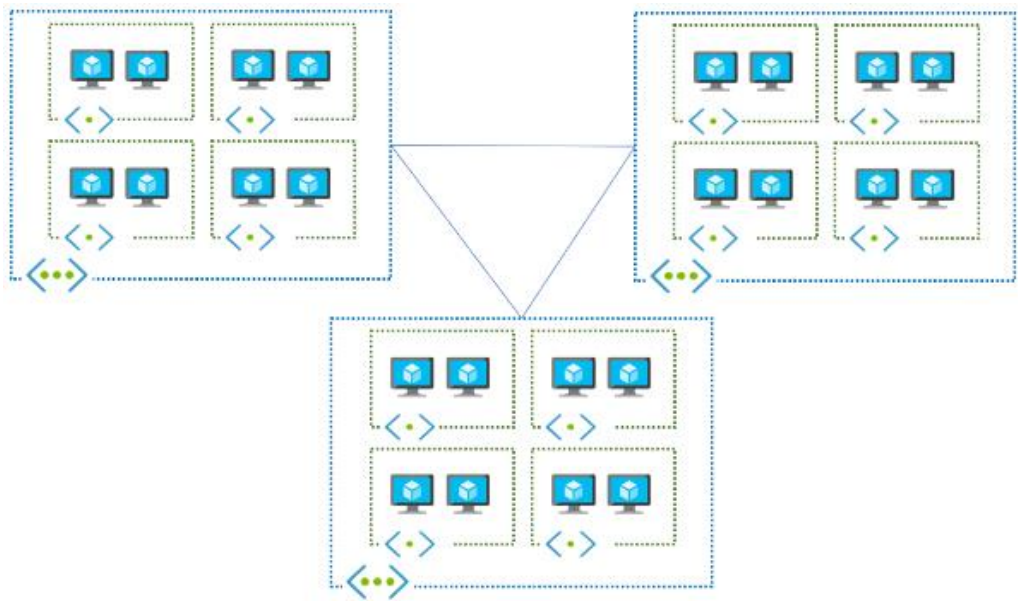

Figure 6. Topology 3 (Azure): further simplified to a spoke full-mesh without centralized firewall

Please consider that the goal of this exercise is not to debate which design is "better", but rather to evaluate which one is "simpler". This additional simplicity would be one of the factors to consider selecting the optimal design, along with others like cost, functionality, etc.

The type-summary graph for this topology without centralized firewall and route tables looks like this:

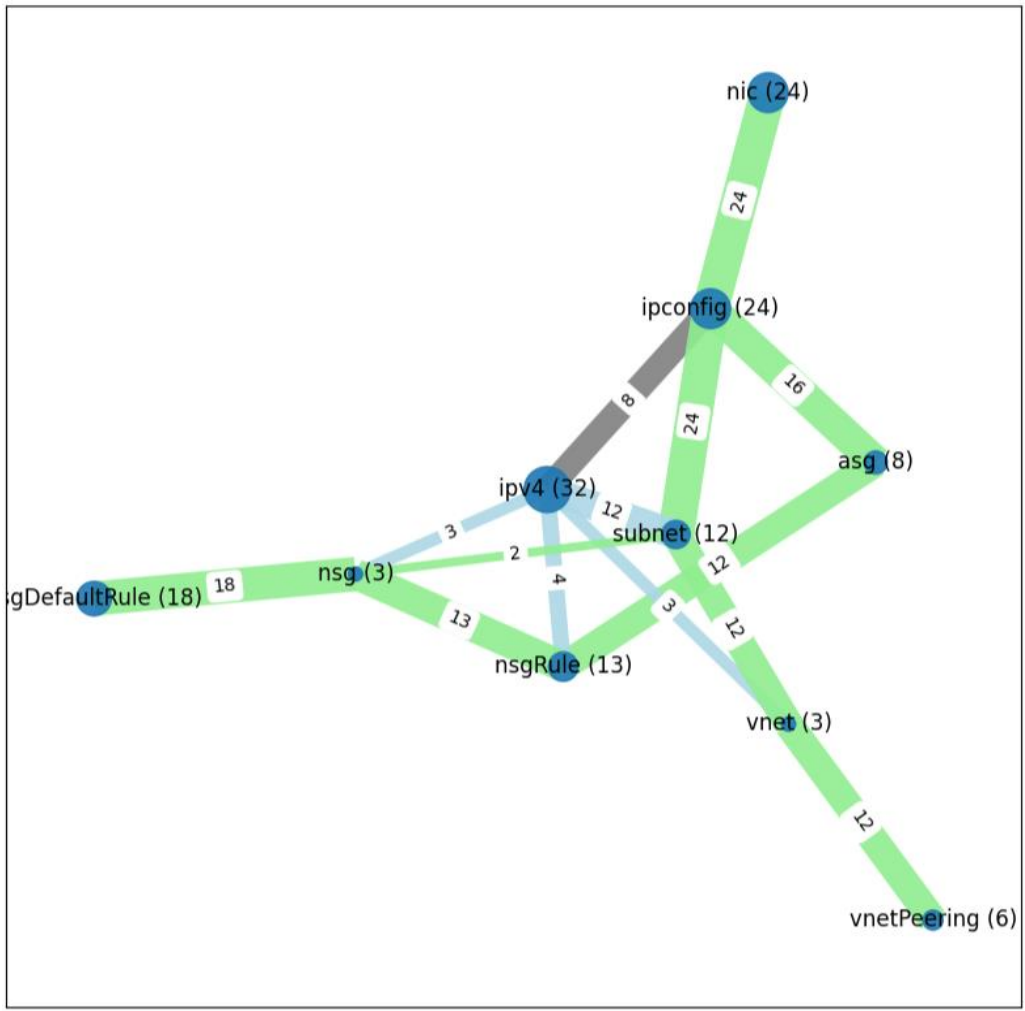

Figure 7. Topology 3 (Azure) – Vertex-type summary graph

Especially in the summary per node types it can be appreciated a lower number of them, what can be verified looking at the metrics for the graph (only 5 I-Types and 4 P-Types, compared to the 8 and 8 of the previous topologies with route tables and a centralized firewall).

The sum of excess degrees of IPv4 nodes is 6 as well, like for Topology 2. While in Topology 2 IP prefixes were used in IP Groups for the firewall policy, in Topology 3 they are used inside of NSG rules. In other words, Topology 3 has traded one complexity for another.

The following table summarizes the key complexity evaluation metrics across all three sample topologies:

Table 3. Topologies 1, 2 and 3 (Azure) – Complexity metrics

| Topology | Nodes/$N_E$ | L-Edges | T-Edges | I-Types | P-Types | IP-ED |
|---|---|---|---|---|---|---|
| Topology 1 (Azure) | 7.33 | 55 | 203 | 8 | 8 | 37 |
| Topology 2 (Azure) | 6.42 | 24 | 163 | 8 | 8 | 6 |
| Topology 3 (Azure) | 5.96 | 22 | 138 | 5 | 4 | 6 |

These three topologies represent two ways of reducing complexity:
- Reduce the number of loose couplings and usage of IP addresses to express policy (from topology 1 to topology 2)
- Reduce the number of nodes (from topology 1 to 2, and 2 to 3)

## 3. COMPARING THE SAME TOPOLOGY ON DIFFERENT ABSTRACTIONS

The second use case of the complexity evaluation framework for network topologies is comparing the same topology across different network abstraction models. The goal is being able to judge the efficiency of different network abstractions related to expressing network policy.

### 3.1. Baseline With Network Command Line

I will take the simplest topology of the previous exercise, and compare it to a baseline. As baseline I have modelled a network configuration run over traditional switches and configured via their command-line interfaces. The node-type summary graph looks like this:

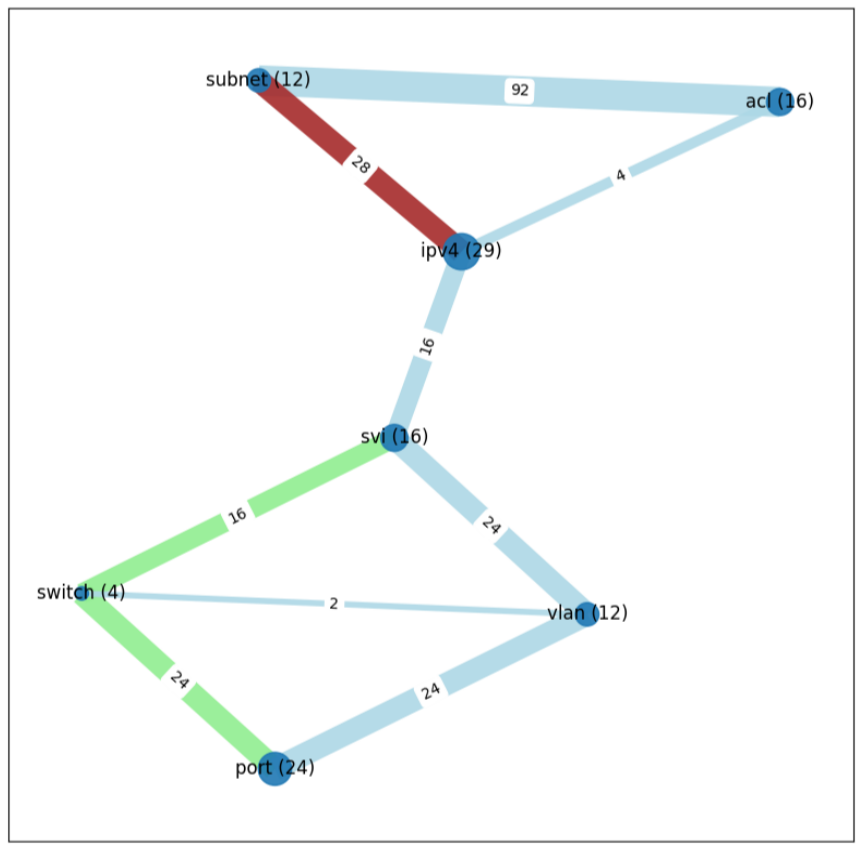

Figure 8. Topology 3 (CLI) – Vertex-type summary graph

The model is very simple (many configurations such as spanning-tree parameters are not expressed in the model, since they are not related to the network policy), and all relationships are loose couplings (except of course the edges between ports and SVIs to individual switches). This is reflected in the graph metrics:

Table 4. Topologies 1, 2 and 3 (Azure) – Complexity metrics

| Topology | Nodes/$N_E$ | L-Edges | T-Edges | I-Types | P-Types | IP-ED |
|---|---|---|---|---|---|---|
| Topology 3 (Azure) | 5.96 | 22 | 138 | 5 | 4 | 6 |
| Topology 3 (CLI) | 4.71 | 186 | 40 | 4 | 1 | 15 |

The CLI model has even fewer nodes than the Azure topology, but it has many more loose couplings. It has as well more infrastructure-related node types, which is not surprising given the fact that the Azure network abstraction doesn't need to care about the underlying infrastructure since it solely the cloud vendor's responsibility.

### 3.2. Kubernetes Network Abstraction

Kubernetes as a solution to deploy application workloads abstracting the underlying hardware has been extremely successful since its inception. Kubernetes was architected around some fundamental principles [14]: its configuration is declarative, the objects in the model are loosely coupled to each other, and the interactions with hardware is abstracted via so-called plugins (such as for networking or storage).

Specifically for the network part, Kubernetes creates a flat layer-3 topology where all workloads are deployed. Optionally, the operator can decide to apply additional security and traffic engineering policies to the network such as network policies or service mesh[15].

The topology that I will model with Kubernetes will be very similar to the previous one deployed on public cloud and CLI, by translating concepts from Azure to Kubernetes when possible, or applying similar ones when not. For example, in Kubernetes there are no virtual networks, VLANs or subnets, all pods receive IP addresses from the same space. However, applications can be separated with namespaces and network policies that refer to labels.

Hence, I have modelled a similar topology consisting of 24 pods (the same number as the virtual machines in the previous examples) and included network policies and services into the configuration. The Kubernetes network configuration will contain three namespaces, each with 4-tier app. Network policies in the first two will only allow communication from one tier to the next, and the last namespace will work as a shared services platform with connectivity to all other workloads:

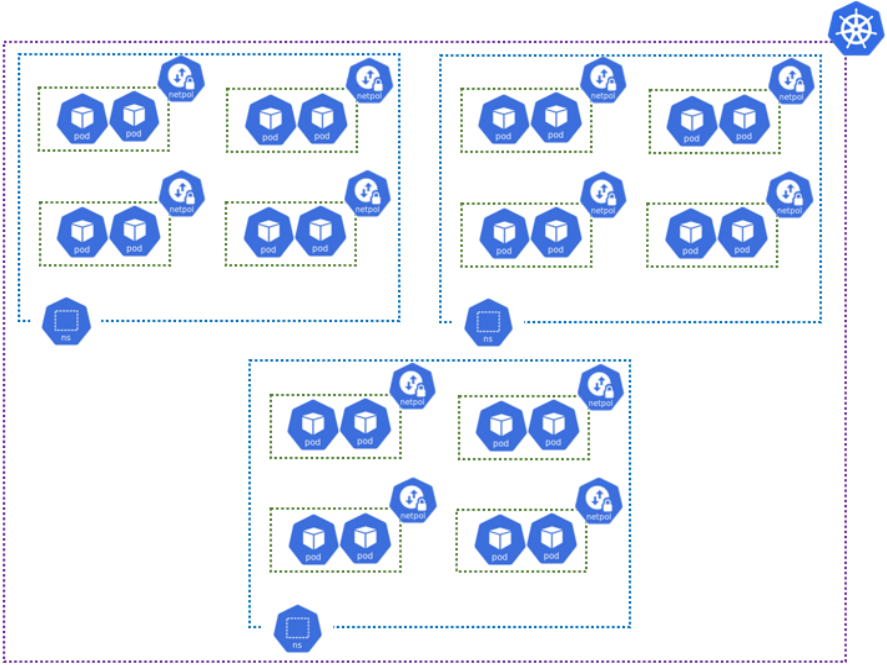

Figure 9. Topology 3 (Kubernetes) – Design

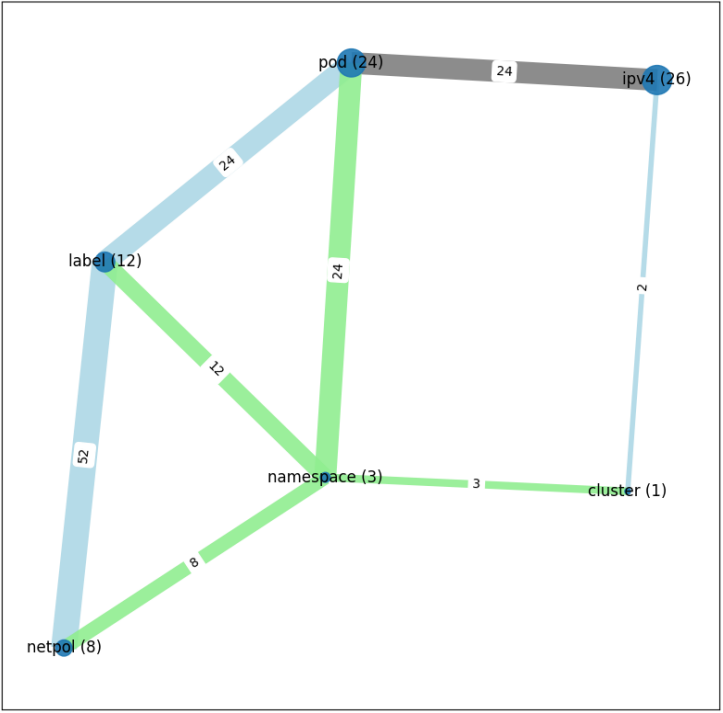

Figure 10. Topology 3 (Kubernetes) – Vertex-type summary graph

The number of vertices is considerably lower than in the previous topologies. The main reason is due to the simplicity of the Kubernetes abstraction, where IP addresses do not play such a prominent role in the configuration. Instead, it is the labels that play a central role replacing other couplings like subnets and virtual networks and acting as the only link between pods and higher-order objects such as network policies and services.

One thing to remark here is that even if the declarative nature of Kubernetes using labels greatly contributes to simplifying the underlying network model, labels are still literal values. As a matter of fact, many Kubernetes misconfigurations can be due to mistyping label names [16].
There is a substantial difference between literals corresponding to IP addresses and labels though: while labels are independent of the topology, IP addresses depend on it. For example, the same deployment across different data centres would probably have different IP addresses, but it could follow the same labels. Hence, there is a qualitative difference between the literals in the Kubernetes network model (labels) and the Azure network model (IP addresses).

Table 5. Topologies 1, 2 and 3 (Azure) – Complexity metrics

| Topology | Nodes/$N_E$ | L-Edges | T-Edges | I-Types | P-Types | IP-ED |
|---|---|---|---|---|---|---|
| Topology 3 (Azure) | 5.96 | 22 | 138 | 5 | 4 | 6 |
| Topology 3 (CLI) | 4.71 | 186 | 40 | 4 | 1 | 15 |
| Topology 3 (K8S) | 3.08 | 82 | 43 | 2 | 3 | 0 |

It is remarkable that the excess-degree of IP nodes is zero in the Kubernetes abstraction. In other words, IP addresses are not used to define policy in Kubernetes, only labels. Another fact to highlight is how few infrastructure-related object types exist in the Kubernetes network model. This is due to the fact that Kubernetes doesn't need concepts like subnets or networks, and it abstracts the network implementation to Container Network Interface (CNI) plugins.

### 3.3. Cisco ACI Network Abstraction

Cisco Application Centric Infrastructure (ACI) is a data centre networking solution that Cisco launched in 2014 [17]. This paper will not try to explain every single ACI concept, but it will focus on a particular aspect: its abstraction to configure networks. Cisco ACI was one of the first network solutions to offer an abstraction to configure it, where the concept of End Point Groups (EPGs) takes the central part. Similar abstractions were used by other networking models such as OpenStack's Group-based Policy [18].

Cisco ACI is a completely different construct than a cloud logical network or Kubernetes, since it needs to provide additional model for the physical infrastructure running beneath. For example, overlay models such as cloud networks or virtual machine hypervisors do not need to bother about how the underlying physical switches are configured, but ACI is that physical switch. And yet, the actual implementation of the underlying data plane is hidden from the user, which is a commonality with Kubernetes and public cloud models.

Consequently, the model in ACI is very different to the topologies we have seen so far. However, the exercise is still valid, and in this example, I will model a very basic ACI topology with 2 leaf switches, the same number of workloads I had in the Azure, CLI and Kubernetes topologies, and endpoint allocation based on physical port (ACI supports other allocation types for virtual workloads, but I will stick to port-based allocation for simplicity).

The following picture describes the overall model, where two 4-tier apps are exposed to a L3out, and there are 4 groups of shared services workloads providing services to each app tier:

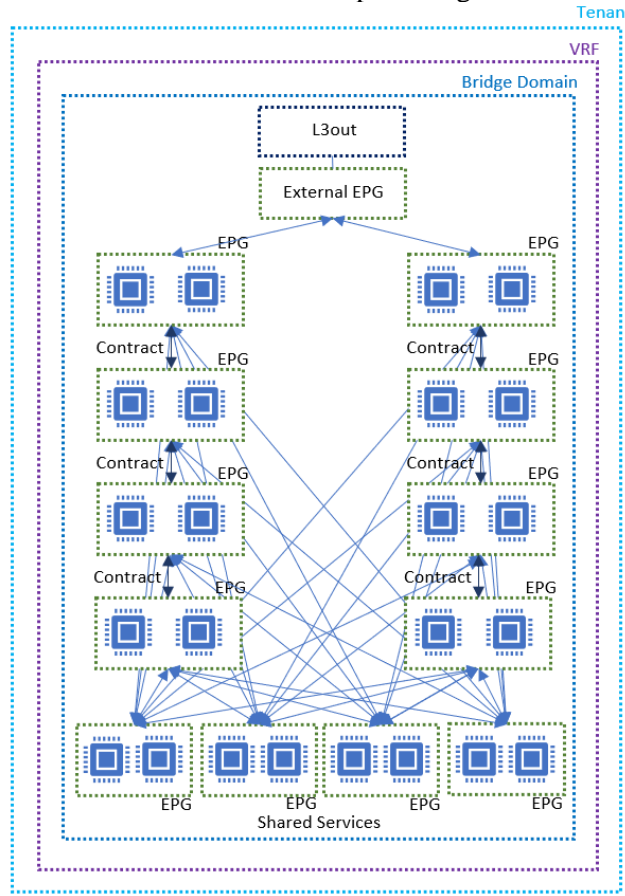

Figure 11. Topology 3 (ACI) – Design

The node-type graph takes this shape:

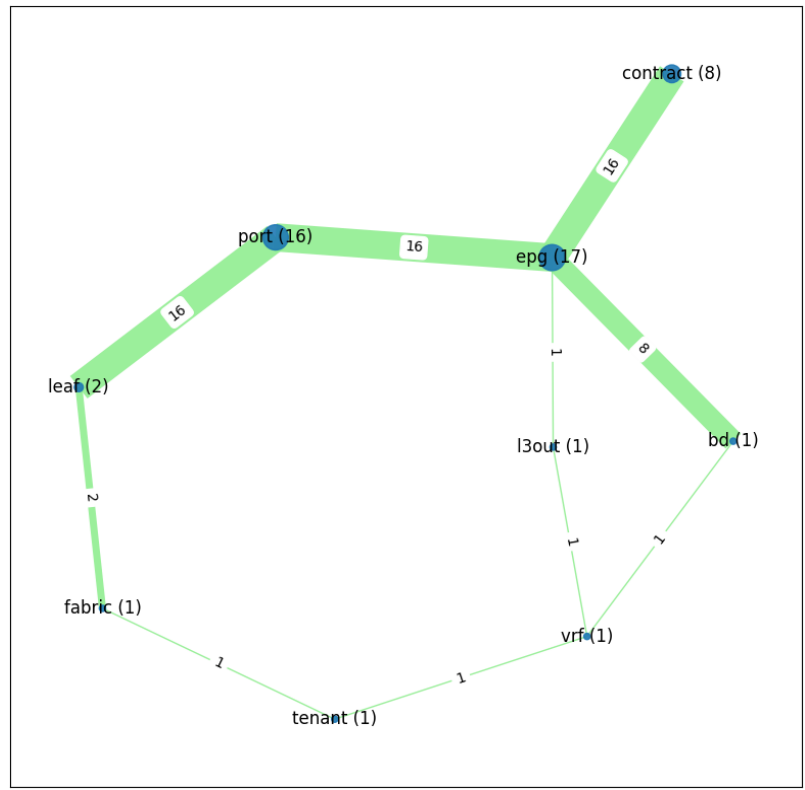

Figure 12. Topology 3 (ACI) – Vertex-type summary graph

The diagram clearly shows the central role that the so-called "End Point Groups" (or EPGs) play in modelling the communication (or lack thereof) between different groups of devices.

As the node-type graph shows, there are other types in the object model such as tenants, VRFs (or contexts), bridge domains, L3 outs, and many more that I haven't included in this exercise, but those are not relevant for policy. For more information on the complete object model of ACI refer to [19].

Table 6. Topologies 1, 2 and 3 (Azure) – Complexity metrics

| Topology | Nodes/$N_E$ | L-Edges | T-Edges | I-Types | P-Types | IP-ED |
|---|---|---|---|---|---|---|
| Topology 3 (Azure) | 5.96 | 22 | 138 | 5 | 4 | 6 |
| Topology 3 (CLI) | 4.71 | 186 | 40 | 4 | 1 | 15 |
| Topology 3 (K8S) | 3.08 | 82 | 43 | 2 | 3 | 0 |
| Topology 3 (ACI) | 2.0 | 0 | 99 | 6 | 3 | 0 |

There are some similarities between the Cisco ACI and the Kubernetes abstractions:
- Both have zero excess degree for IP addresses, meaning that IP addresses don't play a role in the definition of network policy.
- Both have a lean model for policy definition, with only 3 object types (P-Types). The CLImodel is even leaner, but it does so at the cost of using IP addresses to define policy.

The most significant difference between both policy models is the fact that Cisco ACI exclusively leverages tight couplings, while Kubernetes makes a strong use of loosely coupled objects. There are pros and cons of each design, which are out of the scope of this document since their complexity is similar.

## 4. CONCLUSION

Network abstractions are in their infancy. While some vendors such as Cisco with their Application Centric Infrastructure [17] and Juniper with Apstra[20] have innovated in this area, the most widely adopted abstractions have come from public cloud and Kubernetes.

Public cloud network models and abstractions haven't gone all the way they could have, and they are still based on traditional networking concepts such as subnets, which increases the level of infrastructure-type configuration nodes without any value added for the network policy, and hence incur in excessive configuration complexity.

The simplest network abstractions according to the metrics used in this articlewere Kubernetes and Cisco ACI, being public cloud (Azure in this example, but other public clouds follow very similar models) still better than the baseline of command-line interface.

One interesting metric where the Kubernetes model differs from the others is in its loose couplings implemented with free-text labels. Although this mechanism can be prone to errors, it offers a very flexible deployment mechanism, since the different resources can be created (or re-created, when troubleshooting network issues) in no particular order. Although these literal labels mean an increase in complexity, it is worth considering the trade-off.

The implementation of tight couplings is worth a last reflexion:both abstractions evaluated in this article focusing on tight couplings (Cisco ACI and Microsoft Azure) use an identifier derived from the object name as a key for these couplings, which makes the renaming of objects quite problematic. Instead, other public clouds such as AWS use automatically derived GUIDs as key for the couplings. While enabling object renaming, these GUIDs introduce an additional level of complexity for the human operator, who just by looking at a GUID is not able to identify whether the coupling is correct or not).

All in all, the conclusion is that cloud networks should evolve, for example in the direction suggested by [21] or [22]. Regardless of whether the next-generation object model is inspired by Kubernetes (which would add the benefit of the huge ecosystem that has been created in the last years) or more by a group-based policy à la ACI or Apstra, this future network abstraction should possess these attributes:
- IP-address-independent policies for routing and traffic segmentation (low excess degree of IP address nodes)
- Grouping of endpoints based on topology-independent attributes (low number of infrastructure node types)

**ACKNOWLEDGEMENTS**

Many individuals have helped to build this document, but especially Adam Stuart has been instrumental in providing feedback and inspiration.

## REFERENCES

[1] A. Bednarz, "Global Microsoft cloud-service outage traced to rapid BGP router updates," 30 January 2023. [Online]. Available: https://www.networkworld.com/article/3686531/global-microsoft-cloud-service-outage-traced-to-rapid-bgp-router-updates.html.

[2] Uptime Institute, "Outage Analysis," 2022.

[3] A. Bednarz, "Top reasons for network downtime," networkworld.com, 18 November 2016. [Online]. Available: https://www.networkworld.com/article/3142838/top-reasons-for-network-downtime.html.

[4] J. Tantsura and R. White, Navigating Network Complexity, Addison Wesley, 2015.

[5] H. Preston, "What Does "Network as Code" Mean?," 7 February 2018. [Online]. Available: https://blogs.cisco.com/developer/what-does-network-as-code-mean.

[6] A. Leimio, "Rant: Cisco ACI Complexity," 1 March 2021. [Online]. Available: https://blog.ipspace.net/2021/03/rant-cisco-aci-complexity.html.

[7] R. K. Standish, "Complexity of Networks," 2005.

[8] C. A. Huayllaa, M. N. Kuperman and L. A. Garibaldi, "Statistical measures of complexity applied to ecological networks," 2023.

[9] A. Farzaneh, J. P. Coon and M.-A. Badiu, "Kolmogorov Basic Graphs and Their Application in Network Complexity Analysis," Entropy, 2021.

[10] R. Noldus and P. V. Mieghem, "Assortativity in complex networks," Journal of Complex Networks, vol. 3, no. 4, p. 507–542, 2015.

[11] M. Almeida-Neto, P. Guimares, P. R. J. Guimares, R. D. Loyola and W. Ulrich, "A consistent metric for nestedness analysis in ecological systems: reconciling concept and measurement," Oikos, vol. 117, no. 8, pp. 1227-1239, 2008.

[12] R. Solomonoff, "A formal theory of inductive inference," Information and Control, vol. 7, no. 1, pp. 1-22, 1964.

[13] Microsoft, "Azure Virtual Network Routing," 11 February 2023. [Online]. Available: https://learn.microsoft.com/en-us/azure/virtual-network/virtual-networks-udr-overview.

[14] B. Burns, J. Beda and K. Hightower, Kubernetes: Up and Running: Dive into the Future of Infrastructure, O'Reilly, 2019.

[15] Linux Foundation, "Kubernetes Network Policies," 22 December 2022. [Online]. Available: https://kubernetes.io/docs/concepts/services-networking/network-policies/.

[16] T. Khalid, "Common Kubernetes Errors Made by Beginners," 21 July 2021. [Online]. Available: https://medium.com/nerd-for-tech/common-kubernetes-errors-made-by-beginners-274b50e18a01.

[17] J. Ezerski, "Cisco ACI Best Practices: Upgrade your Fabric with Confidence," 26 September 2022. [Online]. Available: https://blogs.cisco.com/datacenter/cisco-aci-best-practices-upgrade-your-fabric-with-confidence.

[18] OpenStack, "Group-based Policy Abstractions for Neutron," 1 July 2014. [Online]. Available: https://specs.openstack.org/openstack/neutron-specs/specs/juno/group-based-policy-abstraction.html.

[19] J. Moreno, F. Dahenhardt and B. Dufresne, Deploying ACI, Cisco Press, 2018.

[20] Juniper Networks, "Juniper Apstra Architecture," May 2021. [Online]. Available: https://www.juniper.net/content/dam/www/assets/white-papers/us/en/2021/juniper-apstra-architecture.pdf.

[21] S. McClure, Z. Medley, D. Bansal, K. Jayaraman, A. Narayanan, J. Padhye, S. Ratnasamy, A. Shaikh and R. Tewari, "Invisinets: Removing Networking from Cloud Networks," usenix, 2023.

[22] M. Dvorkin, "Kubernetes Eats Network," 5 December 2022. [Online]. Available: https://githedgehog.com/kubernetes-eats-network/.

[23] CNCF, "New SlashData report: 5.6 million developers use Kubernetes, an increase of 67% over one year," 20 December 2021. [Online]. Available: https://www.cncf.io/blog/2021/12/20/new-slashdata-report-5-6-million-developers-use-kubernetes-an-increase-of-67-over-one-year/.

[24] J. Song, "Transparent Traffic Intercepting and Routing in the L4 Network of Istio Ambient Mesh," 22 December 2022. [Online]. Available: https://jimmysong.io/en/blog/ambient-mesh-l4-traffic-path/.

[25] C. Hoge, "Advertising Kubernetes Service IPs with Calico and BGP," 1 April 2020. [Online]. Available: https://www.tigera.io/blog/advertising-kubernetes-service-ips-with-calico-and-bgp/.

[26] A. Goryashko, L. Samokhine and P. Bocharov, "About complexity of complex networks," Applied Network Science, no. 4, 2019.

[27] H. Zenil, N. A. Kiani and J. Tegnér, "A Review of Graph and Network Complexity from an Algorithmic Information Perspective," Entropy, 2018.

[28] J. Moreno, "hubandspoke_complex.azcli," 24 April 2023. [Online]. Available: https://github.com/erjosito/azcli/blob/master/hubandspoke_complex.azcli.